# Threshold Resistance in the DC Josephson Effect

Yong-Jihn Kim

*Department of Physics,
University of Puerto Rico,
Mayaguez, PR 00681*

**Abstract.** We show that SIS Josephson junctions have a threshold resistance, above which the Josephson coupling and the supercurrents become extremely small, due to the shrinking of the Cooper pair size during the Josephson tunneling. Accordingly, the threshold resistance is smaller for higher $T_c$ superconductors with small Cooper pair size and for the insulating barrier with higher resistance. This understanding agrees with the observations in SIS junctions of low $T_c$ superconductors, such as Sn, Pb, and Nb. For $MgB_2$ it explains why the big gap does not show the supercurrents, unlike the small gap. Furthermore, it is consistent with the fact that high $T_c$ cuprates show the Josephson effects only for SNS type junctions, including the intrinsic Josephson effects.

**Keywords:** Josephson effects, SIS junctions, Threshold resistance.

**PACS:** 74.50.+r, 74.20Fg, 74.25.Sv

## INTRODUCTION

In 1962 Josephson predicted that supercurrents can flow through the insulating barrier between two superconductors due to the Cooper pair tunneling.[1] The supercurrent, $j$, depends on the relative phase, $\varphi$, of two superconductors, i.e.,

$$j = j_1 \sin\varphi. \qquad (1)$$

The maximum DC supercurrent, $j_1$, was given by[2,3]

$$j_1 = \frac{\pi}{2e}\frac{\Delta}{R_n}, \qquad (2)$$

where $\Delta$ and $R_n$ are the energy gap and the tunneling resistance, respectively.

On the other hand, there are some experiments which show more complicated behavior than Eq. (2) suggests. For instance, the maximum DC supercurrent in low $T_c$ superconductors, such as Pb, Sn, and Nb, decreases much faster than $1/R_n$ above a few ohms.[4,5,6] $MgB_2$ does not show the supercurrent for the big gap of ~7meV.[7,8] High $T_c$ cuprate Josephson junctions are mainly SNS type, including the intrinsic junctions.[9]

We show that SIS Josephson junctions have a threshold resistance, due to the shrinking of the Cooper pair size during the tunneling, which explains the above experiments.

## COOPER PAIR WAVEFUNCTION APPROACH TO JOSEPHSON TUNNELING

We present Cooper pair wavefunction approach to the Josephson effects. It is shown that the Josephson coupling energy, $E_J$, is determined by the overlap of the Cooper pair wavefunctions of two superconductors divided by a thin insulating layer:[10]

$$\begin{aligned}E_J = &V\int F_r^*(x)F_l(x)dx \\ &+V\int F_l^*(x)F_r(x)dx,\end{aligned} \qquad (3)$$

where $F_l$ and $F_r$ are the effective Cooper pair wavefunctions in the left and right sides, and V is the phonon-mediated matrix element. Actually, this equation is closely related to the approximate expression of the supercurrent suggested by Josephson,[1] i.e.,

$$j \cong \frac{1}{2}j_1 F_r^* F_l + \frac{1}{2}j_1 F_l^* F_r. \qquad (4)$$

Note that

$$j = \frac{2e}{\hbar} \frac{\partial E_J}{\partial \varphi}. \quad (5)$$

## THRESHOLD RESISTANCE

It is essential to calculate the tail of the Cooper pair wavefunctions to determine the Josephson coupling energy and the supercurrent. We stress that the Cooper pair size, $\xi_0$, will shrink during the tunneling for the insulating barriers with high tunneling resistance and for high $T_c$ superconductors, including $MgB_2$. This is similar to the reduction of the Cooper pair size due to the impurity potential. We can estimate the threshold resistance, $R_{th}$, above which the shrinking occurs, i.e.,

$$R_{th} \cong Ce^{2\kappa d_{th}}, \quad d_{th} \sim \sqrt{\xi_0 / 2\kappa}, \quad (6)$$

where $\hbar\kappa = \sqrt{2m(U-E)}$, and C is a constant. Accordingly, for the insulator thickness, $d \geq d_{th}$, we find[10]

$$j_1 = \frac{1}{\lambda} \frac{\Delta}{eR_n} \frac{1}{1+2\kappa d} e^{-\frac{1+2\kappa d}{\xi_0}d}, \quad (7)$$

where λ is the BCS coupling constant.

## COMPARISON WITH EXPERIMENTS

The threshold behavior has been found in experiments.[4-8] Figure 1 shows the comparison of our theoretical calculations with experimental results for Sn-SnO-Pb junction (at 1.4K) by Tinkham's group (Ref. 5), Pb-PbO$_x$-Pb junction (at 4.2K) by Schwidtal and Finnegan (Ref. 4), and Nb-NbO$_x$-Pb junction (at 4K) by Octavio's group (Ref. 6). The solid lines are theoretical calculation based on Eq. (7). It is remarkable that above $R_n \sim 40\Omega$ the suppercurrent of Sn-SnO-Pb junction becomes larger than those of Pb-PbO$_x$-Pb and Nb-NbO$_x$-Pb junctions. It is clear that this is due to the existence of the threshold resistance in the SIS Josephson junctions.

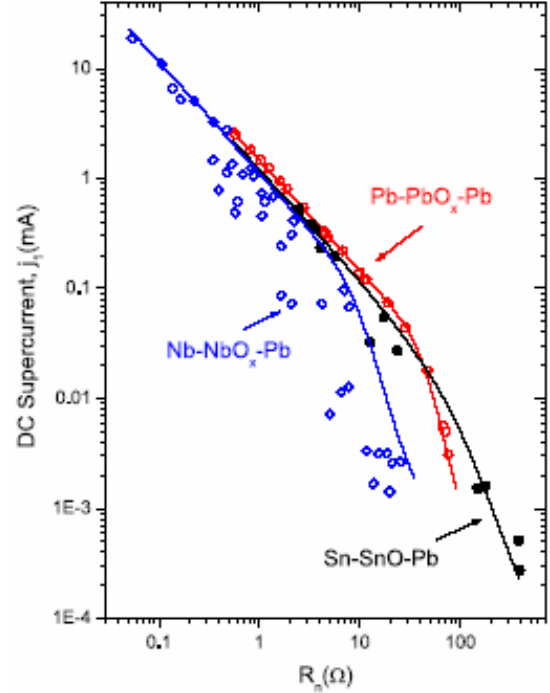

**FIGURE 1.** Maximum DC supercurrent vs tunneling resistance for Sn-SnO-Pb (Ref. 5), Pb-PbO-Pb (Ref. 4) and Nb-NbO-Pb (Ref. 6).

## ACKNOWLEDGMENTS

Supported by the National Science Foundation under Grant No. 0351449.